# ScratchR: sharing user-generated programmable media

Andrés Monroy-Hernández
MIT Media Laboratory
77 Massachusetts Ave. E15-120B
Cambridge, MA 02139 USA
+1 617 253 4191
andresmh@media.mit.edu

**ABSTRACT**
In this paper, I describe a platform for sharing *programmable media* on the web called ScratchR. As the backbone of an on-line community of creative learners, ScratchR will give members access to an audience and inspirational ideas from each other. ScratchR seeks to support different states of participation: from passive consumption to active creation. This platform is being evaluated with a group of middle-school students and a larger community of beta testers.

**Author Keywords**
Programmable media, user-generated content, social networks, creativity, on-line communities, programming languages

**ACM Classification Keywords**
K.3.1 [**Computer Uses in Education**]: Collaborative learning

**INTRODUCTION**
Scratch [1] is a new programming language that makes it easy for youth and children to create programmable media such as animated stories, games and interactive art. ScratchR is the web-based platform that lets kids share their Scratch creations on-line. ScratchR gives kids access to an on-line community of Scratch programmer like themselves. One of the goals of ScratchR is to help the creative process by fostering collaboration. To achieve collaboration, ScratchR tries to provide a wide range of entry points: from the simple act of interacting with a project to commenting and uploading new projects. The platform is a repository of user-generated content that then serves as a source of inspiration and appropriable objects. Lastly, ScratchR allows members to connect with each other, forming a social network of creators. The immediate result of this work is the public release of the Scratch on-line community which is as an important element of the whole Scratch experience. The platform is being evaluated with a small group of middle-school students and a larger group of beta testers. The result of this research will provide a theoretical framework for analyzing and designing on-line communities that engage kids in learning from each other by sharing their creations.

**Programmable media and Scratch**
The easiest way to start describing the term *programmable media* is by giving examples of what it is not. Videos, audio, images and text are not programmable media. Programmable media is created when images, audio and text, are controlled according to certain behavior. Scratch is an authoring environment for creating programmable media. For example, a kid can use the image of a cat (*non-programmable media*) and then define a behavior like "move the cat 10 steps when the right arrow key of the computer is pressed". That mix of behavior and non-programmable media is what I call programmable media.

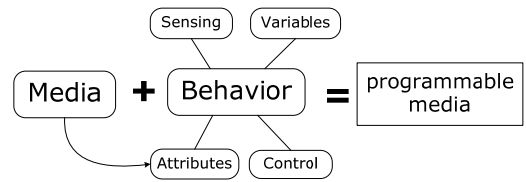

*Figure 1. Elements of a programmable media object.*

**RELATED WORK**
Platforms for sharing user-generated content are not new; in fact, web sites like that have gained a lot of attention in recent years. For example, YouTube [2] and Flickr [3] are well-known platforms for sharing *non-programmable media*. In these sites users get different types of feedback from other members of the community and find inspirational ideas by looking at other people's creations (e.g. pictures and videos). However, there is not a platform for sharing programmable media and especially not one that focuses on youth and children.

**Creative appropriation**
Due to the nature of programmable media, it is very useful to let people reuse other people's work (e.g. blocks of code and sprites). That is what I call *creative appropriation*: the utilization of someone else's creative work in the making of a new object. There are very few platforms that allow creative appropriation; one of them is the OPENSTUDIO [4] web site which focuses on sharing digital paintings. There seems to be no platform that allows people to engage in creative appropriation of programmable media. ScratchR is a unique platform that gives young programmers access to the power of collective intelligence.

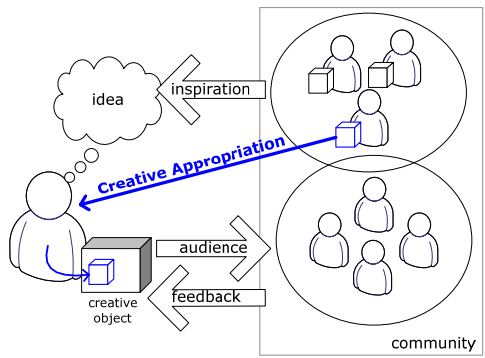

*Figure 2. Platform that fosters creative appropriation*

## DESIGN GUIDELINES

The goal of ScratchR is to foster creative learning as part of an on-line social experience. Inspired by Jenkins' [5] description of the states of participation in fan-fiction communities, I put forward the idea that members of a creative on-line communitie like ScratchR, tend to be in four different states: passive consumption, active consumption, passive production and active production.

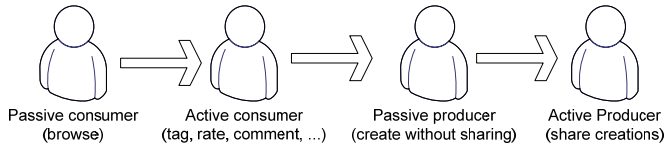

*Figure5. States of participation*

The level of engagement and state of participation varies from person to person and from time to time in each individual. ScratchR is designed to welcome users no matter in what state they wish to be. For example, an active consumer could be engaged in only tagging other people's projects. For active members of the community, be it active consumers or active producers, the social connections and the feeling of being in control of their own community are important characteristics. Overall, the ScratchR platform tries to foster creative learning in a social context by allowing participants to engage in a wide range of activities, such as giving or receiving inspirational ideas and feedback from and to the community.

## ARCHITECTURE

ScratchR is composed of three basic elements: a repository of Scratch projects, a collection of user-generated metadata about those projects and a community of socially networked people. Members of the community can share (upload) or appropriate (download) the original building blocks of the Scratch projects they interact with on the web site. Internally, the platform and the Scratch files themselves keep a record of those activities. Users can also participate in the community by tagging, commenting, grouping and rating other people's projects. These activities occur in the context of a social network where members can connect with each other by establishing friendship links.

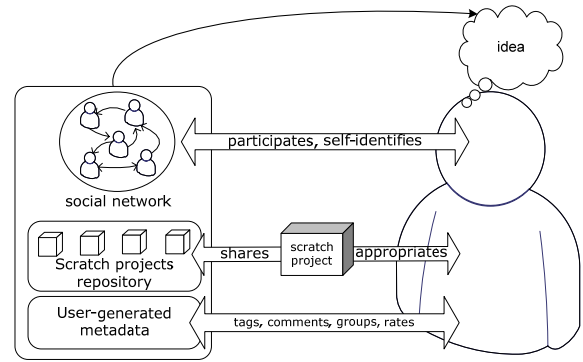

*Figure 3. Components of the ScratchR platform*

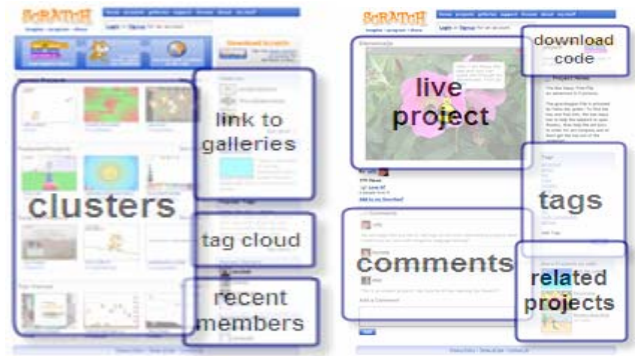

*Figure 4. ScratchR home page and project page*

## EVALUATION

I am working with a group of ten middle-school students, ages 12 to 14, as part of an eleven-week workshop focused on creating programmable media for the web. I expect to collect qualitative information that will help me build illustrative cases. In parallel, quantitative data is being collected as hundreds of children and adults from around the world are using a beta version of the website.

## ACKNOWLEDGMENTS

I would like to thank my advisor Mitchel Resnick and the Lifelong Kindergarten Group for their feedback and support.